\documentclass[twocolumn,preprint,singlespaced]{article}
\usepackage{hyperref}
\usepackage[usenames,dvipsnames]{color}
\hypersetup{colorlinks,citecolor=Blue,linkcolor=Red,urlcolor=Blue}
\usepackage{graphicx}
\usepackage[titletoc]{appendix}
\usepackage{amsmath,amssymb}
\usepackage[T1]{fontenc}
\usepackage[utf8]{inputenc}
\usepackage{authblk}
\usepackage{natbib}
\usepackage[superscript]{cite}

\usepackage[usenames,dvipsnames]{color}
\hypersetup{colorlinks,citecolor=Blue,linkcolor=Red,urlcolor=Blue}
\usepackage{graphicx}

\usepackage{lineno}
\usepackage{amsmath}
\begin{document}

\title{A shorter Archean day-length biases interpretations of the early Earth's climate} 

\author{Christopher Spalding$^1$,$^2$ and Woodward W. Fischer$^1$} 
\affil{$^1$Division of Geological and Planetary Sciences\\
California Institute of Technology, Pasadena, CA 91125} 
\affil{$^2$Department of Astronomy\\
Yale University, New Haven, CT 06511} 
\maketitle 
\begin{abstract}
Earth's earliest sedimentary record contains evidence that surface temperatures were similar to, or perhaps even warmer than modern. In contrast, standard Solar models suggest the Sun was 25\% less luminous at this ancient epoch, implying a cold, frozen planet---all else kept equal. This discrepancy, known as the Faint Young Sun Paradox, remains unresolved. Most proposed solutions invoke high concentrations of greenhouse gases in the early atmosphere to offset for the fainter Sun, though current geological constraints are insufficient to verify or falsify these scenarios. In this work, we examined several simple mechanisms that involve the role played by Earth's spin rate, which was significantly faster during Archean time. This faster spin rate enhances the equator-to-pole temperature gradient, facilitating a warm equator, while maintaining cold poles. Results show that such an enhanced meridional gradient augments the meridional gradient in carbonate deposition, which biases the surviving geological record away from the global mean, toward warmer waters. Moreover, using simple atmospheric models, we found that the faster-spinning Earth was less sensitive to ice-albedo feedbacks, facilitating larger meridional temperature gradients before succumbing to global glaciation. We show that within the faster-spinning regime, the greenhouse warming required to generate an ice-free Earth can differ from that required to generate an Earth with permanent ice caps by the equivalent of 1--2 orders of magnitude of $\mathbf{\textit{p}}$CO$_2$. Accordingly, the resolution of the Faint Young Sun problem depends significantly on whether the early Earth was ever, or even at times, ice-free. 
\end{abstract}

\section{Introduction}

The Earth hosts an extensive variety of habitable environments, which approximately track the places where liquid water persists. Liquid water, therefore, is considered the most critical component of habitability as applied to other planets, ranging from Mars to extrasolar worlds \citep{Kasting2010}. The Earth's geological record has preserved over 4 billion years of global climate history, revealing that large bodies of liquid water have persisted throughout the majority of Earth's past. Accordingly, the Earth is not just habitable today, but has remained habitable for billions of years \citep{Grotzinger1993}. 

Despite the empirically-derived presence of temperate surface environments during Earth's history, models of the Sun's evolution lead to an apparent contradiction. Standard models suggest that the Sun's luminosity has been steadily increasing, such that 3.8 billion years ago, its luminosity was only 75\% of its modern value \citep{Gough1981}. For greenhouse effects similar to today, the planet would be expected to enter a global snowball state at Solar luminosities of only $\sim10\%$ below today's value \citep{Hoffman1998,Yang2012}. In spite of observations to the contrary, the Sun's history predicts a globally glaciated climate during most of the Archean era ($\sim3.8-2.5\,$Gya)\citep{Budyko1969,Sellers1969,Kasting1993,Ikeda1999,Abbot2011}. 

The discrepancy between the presence of liquid water, and a faint early Sun has been dubbed the ``Faint Young Sun Paradox'' and gained widespread awareness owing to the work of \citet{Sagan1972}. A resolution to the paradox requires that the ancient Earth retained a substantially augmented fraction of incoming Solar heat than today, and/or that the reconstructions of Solar luminosity are in error \citep{Feulner2012}. 

A majority of previous work has proposed that the Archean greenhouse atmospheric composition differed radically from today in order to accommodate the fainter Sun \citep{Kasting1993,Wordsworth2013}. Under certain scenarios, climate models possessing heightened levels of greenhouse gases, such as CH$_4$ and CO$_2$, successfully reproduce the required global temperatures under a faint early Sun. Indeed, a promising negative feedback has long been known to exist for generating additional CO$_2$ when the climate cools \citep{Walker1981}. 

Despite the promise shown by the hypothesis of elevated greenhouse warming, geological constraints upon the Archean atmospheric composition are unable to adequately distinguish between paleoclimate models \citep{Feulner2012}. The Archean geological record is sufficiently sparse and fragmented such that it is not clear whether the early Earth was ice-free, or whether ice caps persisted \citep{deWitt2016}. Below we show that distinguishing between these two scenarios is critical to evaluating the likelihood that greenhouse gases can feasibly solve the paradox. Moreover, the paleolatitudes of Archean sedimentary basins are rarely known and hard to measure, making it difficult to ascertain whether or not the geological indicators of a warm climate (e.g. absence of glacial deposits) reflects a truly global condition \citep{Evans2008}.

The early atmospheric composition facilitated the development of life and thus, before settling upon the concentration of greenhouse gases required to explain the geological record, it is important to take full account of climactic drivers aside form atmospheric composition. One such influence is the Earth's rotation rate, which is known to have been decreasing monotonically over geological time owing to tides raised on Earth from the moon \citep{Touma1994}. The most apparent consequence of a faster rotation rate is an enhancement of the equator-to-pole temperature gradient \citep{North1975,Kaspi2015,Liu2017}, which suggests that the Archean Earth exhibited a hotter equator relative to its poles than today. Our results suggest that the extra warmth trapped at the equator generated signals in the geological record that would have tended toward higher temperatures, thereby creating a naturally-biased representation of the mean global Archean climate.

\section{Length of day}

The length of Earth's day was undoubtedly shorter in the past, but an accurate reconstruction back in time is currently beyond model capabilities, largely owing to the uncertain tidal dissipation rates throughout Earth's history \citep{Touma1994,Egbert2000}. Growth bands in the skeletons of marine organisms reveal a Cambrian day length of $\sim21\,$hours \citep{Williams2000}. However, reconstructions of greater antiquity, before abundant availability of animal skeletons, are highly controversial. Despite such limitations, it has been suggested that a 21\,hour resonance in atmospheric thermal tidal forcing maintained a 21\,hour day back as far as 2.5\,Gya \citep{Zahnle1987,Bartlett2016}, and recent stratigraphic techniques are emerging that may soon improve the reliability of Precambrian day length reconstructions \citep{Meyers2018}.

The evolution of the Earth-Moon semi-major axis may be written as \citep{Bills1999}
 \begin{linenomath*}
\begin{align}\label{dissipation}
\frac{da}{dt}=f a^{-11/2}
\end{align}
\end{linenomath*}
where $f$ is a dimensional quantity that encodes information related to Earth's dissipative properties and $a$ is the Earth-Moon semi-major axis (assuming negligible orbital eccentricity). Today, $f\approx6.29\times10^{45}$m$^{13/2}$yr$^{-1}$, measured using the recession of the moon's orbit. Owing to conservation of angular momentum, the Earth-Moon separation is trivially converted to a length of day for the Earth (assuming constancy of structural parameters). A length of day of 6\,hours corresponds to an orbital distance of about 9.26 Earth radii. Such a close orbit would likely leave observable features upon the Earth and so we consider it as a lower limit for the Earth's day-length. 

\begin{figure}[ht!]
\centering
\includegraphics[trim=0cm 0cm 0cm 0cm, clip=true,width=1\columnwidth]{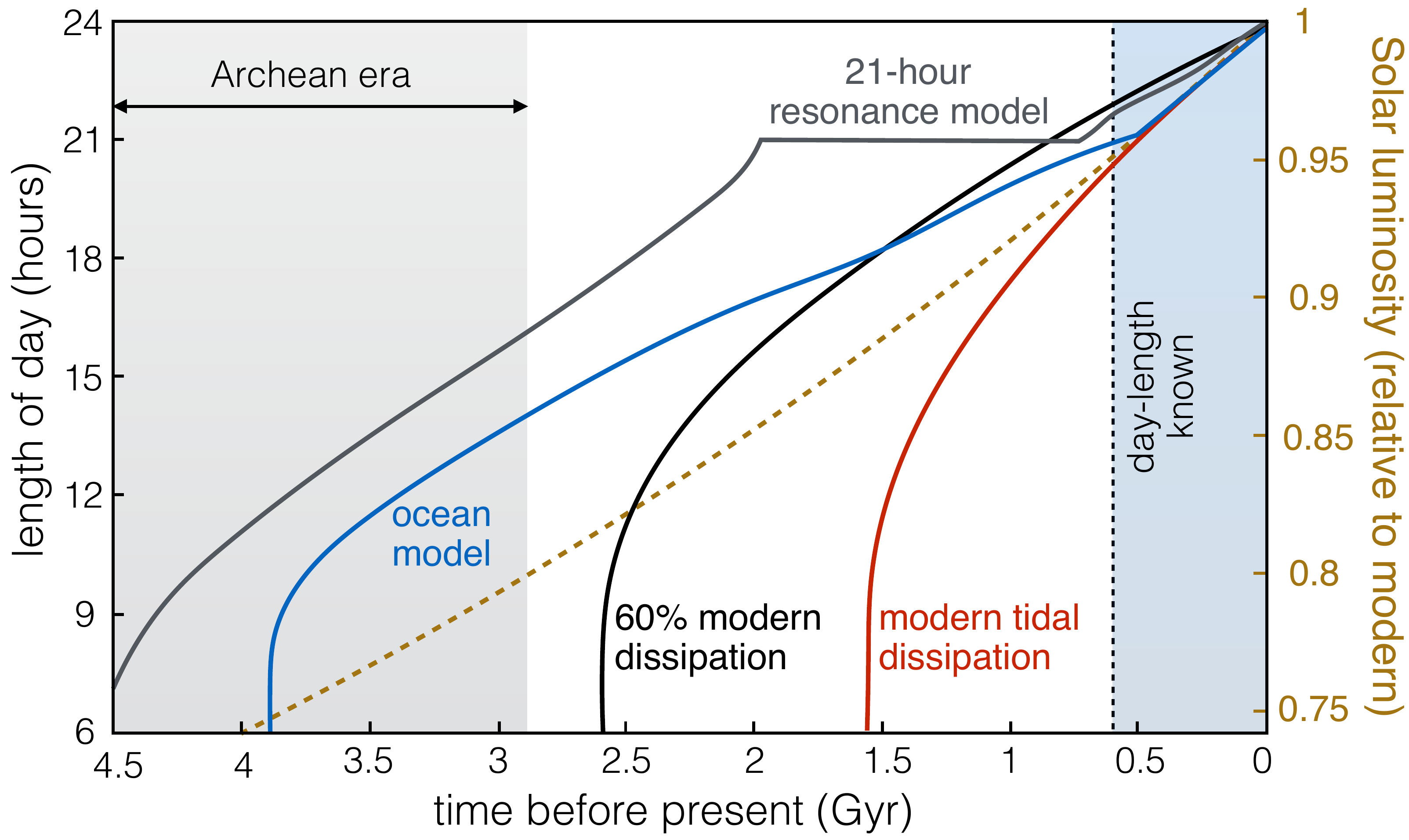}
\caption{The length of Earth's day as a function of time given different assumptions about how tidal dissipation has changed over Earth history. Red, black and gray lines reflect the evolution under modern, 60\% modern and 35\% modern tidal dissipation rates. The blue line, drawn from \citet{Webb1982}, was derived from a model of tidal dissipation with a more sophisticated model of the oceans. All results show that tidal dissipation must be less than 60\% of today averaged over the history of the Earth-moon system, given the lack of evidence for a cataclysmic origin of the moon until prior to $\sim4\,$Gya. By the end of Archean time, the day-length was likely above 12 hours, having changed substantially during that era. The gold line represents the predicted solar luminosity following the computation within \citet{Gough1981}.}
\label{New_Figure}
\end{figure}

The evolution of Earth's length of day is shown in Figure~\ref{New_Figure} for four different possible scenarios, each differing in their assumptions regarding variations in tidal dissipation over time. Assuming a modern dissipation agrees well with geological proxies back to the Cambrian Period \citep{Williams2000,Meyers2018}, but predicts that the Earth and Moon came within 10 Earth radii about 1.5\,Gyr into the past---a condition wholly inconsistent with observations from the Proterozoic and Archean record. Even dissipation 60\% of today's value suggests an inferred age of the Moon that is too young. Assuming 35\% of today's dissipation produces a reasonable age of the moon, but conflicts with the Phanerozoic sedimentary record. 
 
The third model (blue line) was derived by \citet{Webb1982}, and includes a careful computation of the dissipation within an ocean model; it agrees better than a simple modulation of $f$ in equation~\ref{New_Figure}. The fourth hypothesis (shown by a grey line) fits best, which is the 21-hour resonance model as presented in \citet{Bartlett2016}.
 
Given that tides get stronger as the moon is closer, the day length would increase at a faster rate during the Archean than during later time periods, such that the early and late Archean worlds would differ markedly in their rotation rates, perhaps by a factor of 2 (Figure~\ref{New_Figure}). The time spent between day-lengths of 6 and 9 hours is geologically brief, such that for most of the Archean, the Earth was likely rotating at a rate above 6 hours. 

 In order to take account of the range of possibilities, we investigated spin rates $\Omega$ ranging from modern to the equivalent of 6 hours, i.e., $1\leq\Omega/\Omega_0\leq4$, and $\Omega_0$ denotes the modern-day spin angular velocity. This choice of range is in part owing to recent General Circulation Model (GCM) simulations that performed numerical experiments across this 4-fold range of spin rates \citep{Liu2017}, allowing us to calibrate our simplified mathematical framework using results of more complex models. However, the 6\,hour case should be seen as a lower limit to the day length---most of Archean time was likely characterized by a slower spin rate, between 9 and 18 hours (Figure~\ref{New_Figure}).

\section{Influence of Earth's spin rate}

Earth's climate is driven by the distribution of Solar insolation, with more energy being received at the equator than at the poles. Owing to the presence of the atmosphere, heat is transported from hotter to cooler latitudes via turbulent processes that may be modelled as an effective diffusivity, which lowers the meridional temperature gradient. We write this diffusivity as (see appendix)
\begin{linenomath*}
\begin{align}\label{diffusivity}
D\equiv \frac{c_p \Phi}{g R^2}K_{\textrm{eff}},
\end{align}
\end{linenomath*}
where $c_p=10^3$J\,kg$^{-1}$K$^{-1}$ is the specific heat of air, $\Phi=1$\,bar is the surface air pressure, $g=10$\,m\,s$^{-2}$ is the acceleration due to gravity and $R=6400$\,km is the Earth's radius. $K_{\textrm{eff}}$ represents the efficiency of meridional heat transfer by way of turbulent eddies \citep{North1975}, and is the parameter most directly influenced by rotation rate, as we discuss below.

Equation~\ref{diffusivity} indicates that meridional heat transport depends not just upon spin rate, but also upon atmospheric pressure and heat capacity \citep{Williams1997}. Some evidence has been put forward that the atmospheric pressure differed in the past \citep{Goldblatt2009}, though theory and geological proxies often disagree over whether the pressure was more or less than today. If the atmosphere was thick with CO$_2$, its heat capacity may have been larger, additionally helping to warm and homogenize the global climate. However, preserved raindrop imprints, and the size distribution of basaltic vesicles \citep{Som2016}, suggest atmospheric pressures similar to, or lower than, modern. Given these discrepancies, we focus upon the influence of spin rate, but note that a full understanding of the ancient climate requires an auxiliary treatment of the interplay between, as yet uncertain, changes in atmospheric composition, alongside eddy-driven heat transport.

Generally, the magnitude of $K_{\textrm{eff}}$ is reduced at faster rotation rates \citep{Liu2017}, such that in order to retain the same equator-to-pole heat flux, a steeper temperature gradient is retained. The value of $K_{\textrm{eff}}$ cannot be derived from first principles, and must either be measured directly, or extracted from General Circulation Models. In this work, we chose values of $K_{\textrm{eff}}$ as inferred from the simulations performed in \citet{Liu2017}. We fitted the dependence of diffusivity and spin rate from their simulations with the relationship 
 \begin{linenomath*}
 \begin{align}
 \bigg(\frac{K_{\textrm{eff}}}{K_0}\bigg)=\bigg(\frac{\Omega}{\Omega_0}\bigg)^{-n},
 \end{align}
  \end{linenomath*}
 where $n=0.7$ and $K_0\equiv1.7\times10^6\textrm{m}^2\,\textrm{s}^{-1}$.
 
The precise magnitude of $K_{\textrm{eff}}$ varies between investigations. The important feature here is the relationship between $\Omega$ and $K_{\textrm{eff}}$, which we fixed at $K_{\textrm{eff}}\propto \Omega^{-0.7}$. However, it is interesting to compare this to dependences derived elsewhere. \citet{Kaspi2015} performed global climate simulations at various spin rates within the context of extrasolar planets. They did not explicitly measure $K_{\textrm{eff}}$, but the heat flux from eddies $F_E$ was presented. In a diffusive parameterization, $F_E\propto K_{\textrm{eff}}/\Delta T$, where $\Delta T$ is the equator-to-pole temperature difference. Reading $F_E$ and $\Delta T$ from their figure~8 yields an approximate relationship of $K_{\textrm{eff}}\propto\Omega^{-0.71}$, which is remarkably similar to that used here, as derived from \citet{Liu2017}. However, the earlier work of \citet{Williams1997} utilized a relationship where $n=2$: a much steeper relationship. Accordingly, the dependence upon spin rate may be even greater than that derived here, though the qualitative picture will not change.

A rise in meridional temperature gradient is a robust outcome of global climate models incorporating faster spin rates \citep{Kaspi2015,Liu2017}. However, an alternative approach used in previous work is to utilize the principle of maximized entropy production to derive meridional temperature gradients \citep{Lorenz2001}. Though this approach was met with moderate success for Titan and Mars, the methodology has been criticized, and its utility has been questioned \citep{Goody2007}. Thus, we continue with the assumption that rotation rate plays an important role in heat transport via the atmosphere, owing to the extensive and consistent literature surrounding these mechanics.

More uncertain than the relationship between meridional temperature gradient and spin rate, is the influence of spin upon global mean temperature. As the planet spins faster, the Hadley cell is reduced in meridional extent \citep{Kaspi2015}, which potentially reduces cloud coverage and therefore albedo. Consequently, more Solar insolation is absorbed and the planet warms \citep{Liu2017}. This argument is encouraging for the faint young Sun problem, however, it is important to note that different global climate models have exhibited contradictory results regarding the impact of spin rate upon global cloud cover and global temperature \citep{Vallis2009}. Thus we focus here predominantly upon the meridional temperature profile and the ice-albedo feedback, rather than upon the global mean temperature.

In what follows, we argue that the enhanced meridional temperature gradient favors the deposition of sedimentary rocks at low latitudes, skewing the Archean record toward warmer temperatures. We identified that rotation rate and resulting temperature distribution, have two primary influences on interpretations of the ancient climate record. First, the most basic interpretation of the Archean record is that the planet was not globally glaciated. We discuss the role of a faster rotation rate in decreasing sensitivity of the planet to the ice-albedo feedback. Second, we discuss the tendency of a faster-rotating temperature distribution to have biased the deposition pattern of Archean sedimentary rocks toward warmer waters, and in turn, biasing interpretations of the ancient climate.

\section{Spin rate and the ice-albedo feedback}

In order to analyze the influence of rotation rate upon the planet's susceptibility to the ice-albedo feedback, we constructed a simple energy balance model based upon the Budyko-Sellers framework \citep{Budyko1969,Sellers1969,North1975}. A characteristic of this theoretical approach is that within a specific range of Solar luminosities, multiple stable solutions exist. These include an ice-covered state, an ice-free state, along with a solution possessing ice extending from the poles to an ``ice-line'' latitude denoted $\lambda_s$. The value of $\lambda_s$ was computed using a predefined critical temperature for permanent ice caps ($T=T_s\equiv-10^\circ$C). The simplicity of this model allows a mechanistic exploration of the relationship between spin rate and the ice-albedo feedback. 

\begin{figure}[ht!]
\centering
\includegraphics[trim=0cm 0cm 0cm 0cm, clip=true,width=1\columnwidth]{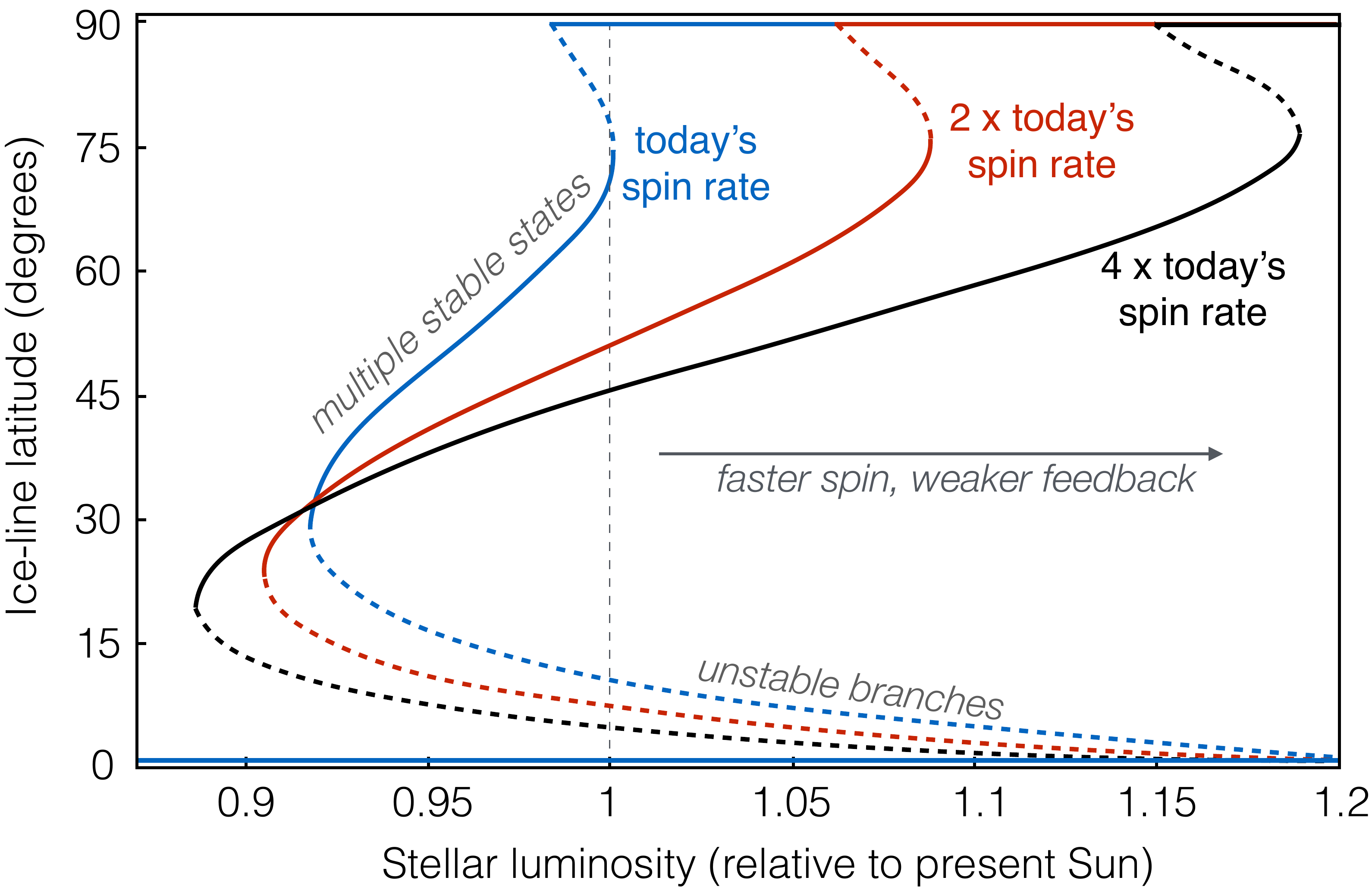}
\caption{The equilibrium ice line (latitude at which $T=-10^\circ$C) as a function of Solar luminosity, for 3 spin rates. The solution is stable if the slope is positive (solid lines) and unstable if the slope is negative (dashed lines). The reduced slope within the stable regions for higher spin rates amounts to a greater resilience to the ice-albedo feedback.}
\label{Hysteresis}
\end{figure}

 The zonally-averaged temperature at any given latitude was computed as a balance between incoming, short-wave radiation (from the Sun), out-going longwave radiation (from the Earth and its atmosphere), together with the effective diffusion of heat between latitudes, with efficiency $K_{\textrm{eff}}$ (defined in section 2). The diffusive term represents turbulent phenomena that tend to transport heat from the equator to the poles, such as baroclinic instabilities \citep{Kaspi2015}.
 \begin{table}
  \centering
\begin{tabular}{ |p{0.5cm}|p{0.5cm}|p{1.1cm}|p{1.5cm}|p{2.3cm}|  }
 \hline
 \multicolumn{5}{|c|}{Parameter values} \\
 \hline
$\,\,\,a_0$ & $\,\,\,a_1$& $\,\,\,\,\,S_2$&$A$\,\,(Wm$^{-2}$) &$B$\,\,(Wm$^{-2}$K$^{-1}$) \\
0.68  & 0.38 & $-0.482$ & $\,\,\,\,196.2$  & $\,\,\,\,\,\,\,\,\,1.953$\\
 \hline
\end{tabular}  
 \caption{The parameter values chosen to solve the Budyko-Sellers type energy balance model. Values taken from \citep{North1975}, except $A$ which was tuned to match today's climate with with $K_{\textrm{eff}}=1.7\times10^{6}$m$^2$s$^{-1}$}
  \label{Param}
\end{table}

Table~\ref{Param} contains the values for the various parameters used, which were largely taken from \citet{North1975}. In Figure~\ref{Hysteresis} we plotted the solution for the ice line as a function of Solar luminosity for 3 values of spin rate, translating to 3 values of $K_{\textrm{eff}}$. The modern climate possesses ice caps down to roughly 72$^\circ$ of latitude, and this was used to tune the parameters for the blue line, representing today's spin rate \citep{North1975}. The other two lines represent spin rates of $2\Omega_0$ and $4\Omega_0$. 
 \begin{figure}
\centering
\includegraphics[trim=0cm 0cm 0cm 0cm, clip=true,width=1\columnwidth]{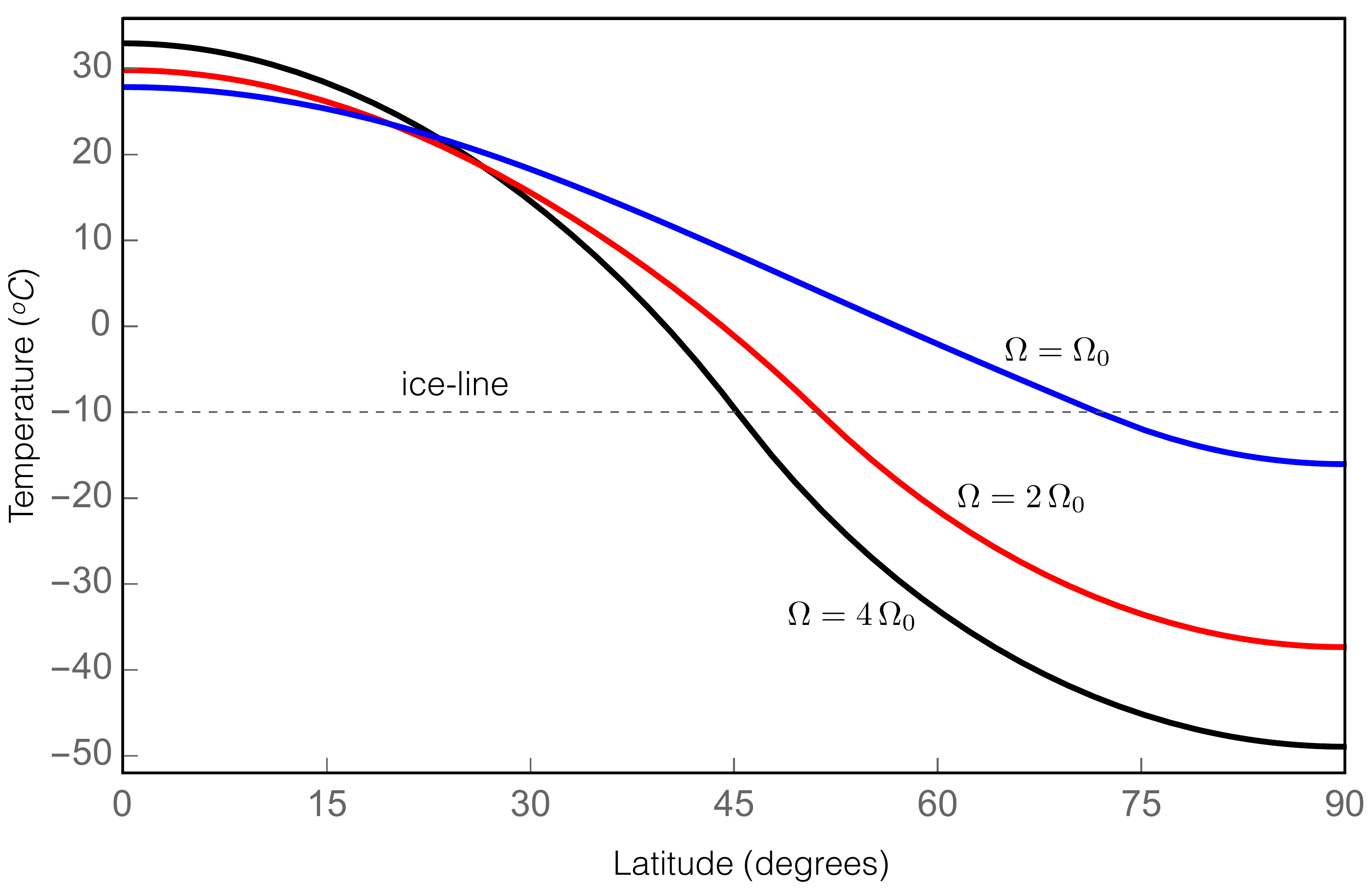}
\caption{Meridional temperature profile arising from the Budkyo-Sellers model for 3 spin rates. Upon comparison to Figure 1, the presence of ice is capable of decreasing polar temperatures while maintaining a warm equator. Faster spins enhance this equator-to-pole difference.}
\label{IA_T}
\end{figure}

Although the earliest descriptions of the faint young Sun paradox were motivated by raising global mean temperature, more recent treatments outline the importance the Earth's susceptibility to global glaciations \citep{Kasting1993,Ikeda1999}. A drop in Solar insolation leads to the spread of ice, which causes an increase in albedo. In reflecting extra sunlight, this additional albedo amplifies the initial drop in insolation. Figure~\ref{Hysteresis} indicates that the faster the spin rate, the weaker the feedback. 

The aforementioned influence upon the feedback strengths lead to two critical and complementary findings. The first, is that ice may extend further toward the equator without triggering a global snowball, providing a greater chance that the equator may be kept ice-free despite low Solar luminosity. However, a second, larger effect is that a lower minimum luminosity is required to entirely deglaciate the planet from a glaciated starting point. This impacts solutions to the faint young Sun paradox because there are substantial differences based on whether or not the early Earth was ice free (even at the poles). Figure~\ref{Hysteresis} captures the result that an ice-free condition is significantly more difficult to achieve under faster spin rates.

\section{Depositional biases}

The inference of a warm Archean Earth is based largely upon evidence from sedimentary rocks associated with early cratons and marine platforms \citep{Grotzinger1993}. Though controversial, absolute temperature estimates have been attempted using these records, from which the overlying marine waters appear to have been at least as warm as surface seawater today \citep{Knauth2003}. However, a key unknown with respect to these early archives is the latitude, and often the environment, of deposition. In this section, we examine how the lithotypes used to infer Archean climate are typically biased toward sedimentation in warmer waters, thereby skewing the implied Archean climate toward warmer temperatures. In conjunction with the enhanced equator-to-pole temperature gradient associated with a faster rotation rate, Archean climate may have been significantly cooler than implied by a naive reading of the geochemical record.

\begin{figure}[h!]
\centering
\includegraphics[trim=0cm 0cm 0cm 0cm, clip=true,width=1\columnwidth]{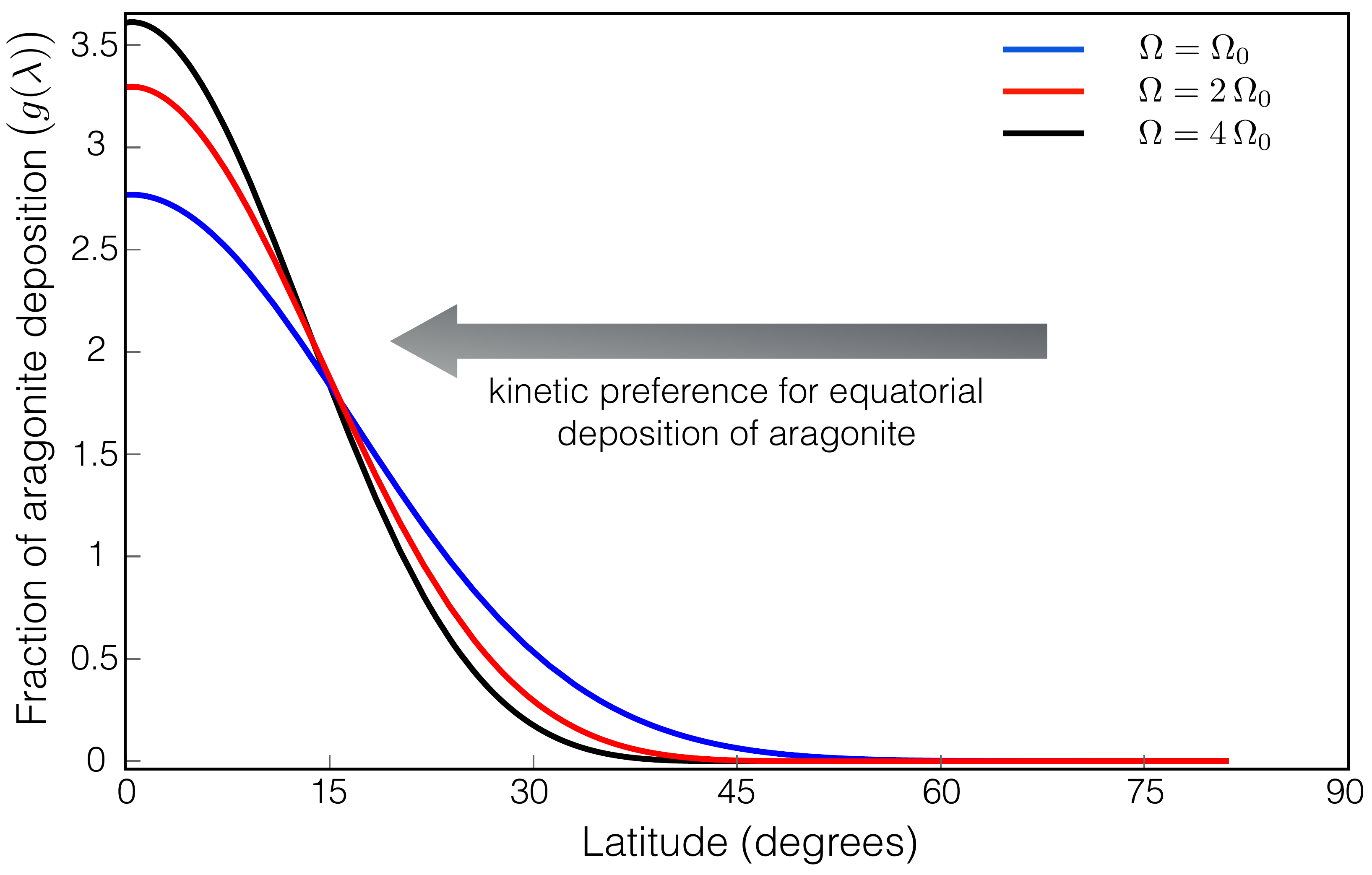}
\caption{The distribution of carbonate deposition rates as a function of latitude for 3 spin rate scenarios (modern; blue, twice modern; red, and four times modern; black) for the Budyko-Sellers climate models presented in the text. All curves are normalized to have an area under the curve of unity, such as to highlight the meridional distribution of deposition, as opposed to the absolute magnitude, which would depend more sensitively upon whole ocean carbonate chemistry. Notice that, particularly when ice occurs at the poles, the rate of aragonite deposition has decreased by an order of magnitude by $\sim30-40^\circ$ and has essentially dropped to zero by $\sim60^\circ$.}
\label{latitudeOmega}
\end{figure}
 
 \subsection{Meridional bias in carbonate deposition}

Some of the most reliable evidence for a temperate Archean climate stems from the existence of shallow marine carbonate platforms \citep{Sumner1996}. In addition, widespread measurements of stable oxygen isotope ratio data from carbonates and cherts have have been interpreted as indicative of high temperatures \citep{Knauth2003}. Accordingly, inferences regarding the paleoenvironment of the Archean by way of sedimentary records are filtered through the lens of geochemical environments favoring the deposition of carbonates and cherts.

In modern oceans, calcium carbonate salts are preferentially deposited in warm, tropical waters. This pattern reflects the evolution and ecology of organisms with carbonate skeletons, but it is also directly related to the inorganic chemistry of the carbonate system \citep{Zeebe2001}, whereby warmer waters are more saturated with respect to CaCO$_3$ salts. The dominant mechanisms of carbonate precipitation during Precambrian time remain uncertain \citep{Grotzinger1983,Grotzinger2000,Ridgwell2003,Higgins2009,Fischer2009}.

Key differences with modern times include a lack of skeletonized metazoa and open-ocean planktic calcifiers. In such a global regime, abiotic mechanisms of carbonate production would have dominated the carbon cycle; fluxes of dissolved inorganic carbon and alkalinity into the oceans would be balanced almost entirely by abiotic precipitation \citep{Ridgwell2003}. Moreover, the predominance of alkalinity-favoring, anaerobic respiration in shallow sediments would amplify carbonate precipitation and preservation within sediments \citep{Higgins2009}.

In Archean basins, carbonate precipitation would have predominantly occurred in shallow-water shelf environments, with a spatial distribution governed largely by temperature. Inferences regarding global climate by way of Archean carbonates must therefore acknowledge the existence of a strong chemical preference, both kinetically and thermodynamically, for precipitation of these rocks within the warmest shallow water environments on the planet. In what follows, we estimate the expected bias associated with a rotationally-enhanced equatorial temperature gradient, and argue that the widespread occurrence of liquid water observed may simply be reflecting a warm equator, kept stable by the fast-rotating planet.

\subsection{Kinetic model of Carbonate deposition}

 The saturation state of seawater with respect to aragonite $\Omega_A$ decreases toward the poles in the modern oceans \citep{Jiang2015}. This trend generally arises from two causes: 1) aragonite is more soluble in colder waters, and 2) the carbonate system adjusts in the presence of an atmosphere containing CO$_2$ to raise the concentration of Dissolved Inorganic Carbon (DIC) in colder waters. In today's oceans, the saturation state of aragonite varies with temperature approximately linearly as
\begin{linenomath*}
\begin{align}\label{latOmega}
\Omega_A\approx a(T/^\circ\textrm{C})+b.
\end{align}
\end{linenomath*}
where $a\approx0.09$ and $b\approx 1.5$ in the modern ocean \citep{Jiang2015}. 

Carbonate compensation is broadly considered to have been limited by $\Omega_A$ over geological time \citep{Grotzinger2000,Ridgwell2003}, as opposed to Ca$^{2+}$-limitation. Accordingly, no matter the precise dependence of $\Omega_A$ upon temperature and latitude, the Precambrian carbon cycle is expected to have exhibited a drop in $\Omega_A$ from the equator toward the poles.

To proceed, we adopted this relationship for the modern, but note that carbonate deposition in Precambrian oceans may not have exhibited a linear dependence. The values of $a$ (the meridional gradient) and $b$ (the saturation state at the freezing point) are also likely to change, but not independently. In particular, if $a$ is raised, in order to achieve an equivalent global precipitation of carbonate, $b$ must drop. Physically, if saturation state varies more markedly with latitude, the saturation state at high latitudes may be lower while retaining an equivalent global deposition. A stronger or weaker dependence will constitute an additional determinant upon potential biases that we do not tackle here. Our goal is simply to compare the biases associated with different rotation rates, for which it suffices to approximate the dependence as linear across all rotation rates.

We used the dependence given by equation~\ref{latOmega} to construct an idealized kinetic model of carbonate deposition. The precipitation rate per unit area of calcium carbonate is generally written \citep{Zeebe2001}
\begin{linenomath*}
\begin{align}
r(\lambda)=k_0\exp(-E_a/R_gT(\lambda))(\Omega_A(T(\lambda))-1)^{n_0},
\end{align}
\end{linenomath*}
where $k_0$ is a reference rate, $\Omega_A$ is the saturation state of aragonite, $n_0\approx1.7$ \citep{Romanek2011} is the order of the reaction, $E_a\approx71$\,kJ\,mol$^{-1}$ is the activation energy and $R_g=8.31$J\,mol$^{-1}$K$^{-1}$ is the universal gas constant. 

The rate of the reaction depends upon temperature in two fundamentally different aspects. The first is that the saturation state itself varies with latitude, being larger in warmer tropical waters. However, the dominant effect lies in the Arrhenius factor $\exp(-E_a/R_gT)$, such that if the whole ocean is to balance the fluxes of ions entering it, calcium carbonate will precipitate faster in the warmer regions owing to a kinetic enhancement of reaction rates. 

To illustrate the effect of temperature, we defined the relative rate of aragonite deposition within a latitudinal band $d \lambda$ as $g(\lambda)d\lambda$, where,
\begin{linenomath*}
\begin{align}
g(\lambda)=r(\lambda) \cos(\lambda),
\end{align}
\end{linenomath*}
which follows from the definition of $r$ as a rate per unit area on the globe. The absolute magnitude of $k_0$ is unimportant for our investigation; what matters is the relative precipitation rate at each latitude. Accordingly, we treated $k_0$ as a normalization factor that satisfies
\begin{linenomath*}
\begin{align}
\Bigg[\int_0^{\pi/2}g(\lambda)d\lambda\Bigg]=1.
\end{align}
\end{linenomath*}

Figure~\ref{latitudeOmega} illustrates the relative degree of carbonate precipitation as a function of latitude. The temperature $T(\lambda)$ was taken from the Budyko-Sellers temperature model (Figure~\ref{IA_T}), using an equivalent solar forcing of the modern day. Note that no deposition occurs at latitudes that are ice-covered year-round. Latitudes with annual mean temperatures below freezing, but above $T_s=-10\,^\circ$C will allow deposition during the fraction of the year when ice is not present. Though strictly unphysical, the rate at these latitudes is reflected by allowing for sub-freezing temperatures in the computation of $g(\lambda)$, rather than fixing the temperature at the freezing point. The two choices do not significantly differ, given the slow precipitation rates at such low temperatures relative to the equator.

\subsection{Preference for equatorial carbonate deposition}

 As is illustrated in Figure~\ref{latitudeOmega}, a significant enhancement of deposition rates near the equator occurs as a result of the combination of enhanced temperatures and surface area at low latitudes. An order of magnitude enhancement of carbonate deposition occurs at the equator relative to the lower latitudes of $\sim30-40^\circ$, and deposition drops to zero close to the ice-line. Crucially, the existence of systematic bias is present for all 3 rotation rates, meaning that the Precambrian climate record exhibits an ever-present tendency towards an equatorial bias.
 
Despite the presence of bias irrespective of rotation rate, the magnitude and interpretation of the bias depends upon rotation rate. Upon comparison to Figures~\ref{Hysteresis} and~\ref{IA_T}, at 1 Solar luminosity, ice extends significantly further toward the equator at higher rotation rates than modern. In other words, receiving information only about lower latitudes gives a significantly more biased picture of global mean temperatures when rotation rates are higher. In order to quantify this bias, we defined the true mean global temperature as
\begin{linenomath*}
\begin{align}
\bar{T}&\equiv\frac{\int_{0}^{\pi/2}T(\lambda)\cos(\lambda)d\lambda}{\int_{0}^{\pi/2}\cos(\lambda)d\lambda}\nonumber\\
&=\int_{0}^{\pi/2}T(\lambda)\cos(\lambda)d\lambda.
\end{align}
\end{linenomath*}
In contrast, the averaged signal captured by paleoclimate records in the geological record will be skewed by the kinetic factor $g(\lambda)$, such that the mean inferred from the record is given by
 \begin{linenomath*}
\begin{align}
\bar{T}_{obs}=\int_{0}^{\pi/2}g(\lambda)T(\lambda)d\lambda.
\end{align}
\end{linenomath*}

\begin{figure}
\centering
\includegraphics[trim=0cm 0cm 0cm 0cm, clip=true,width=1\columnwidth]{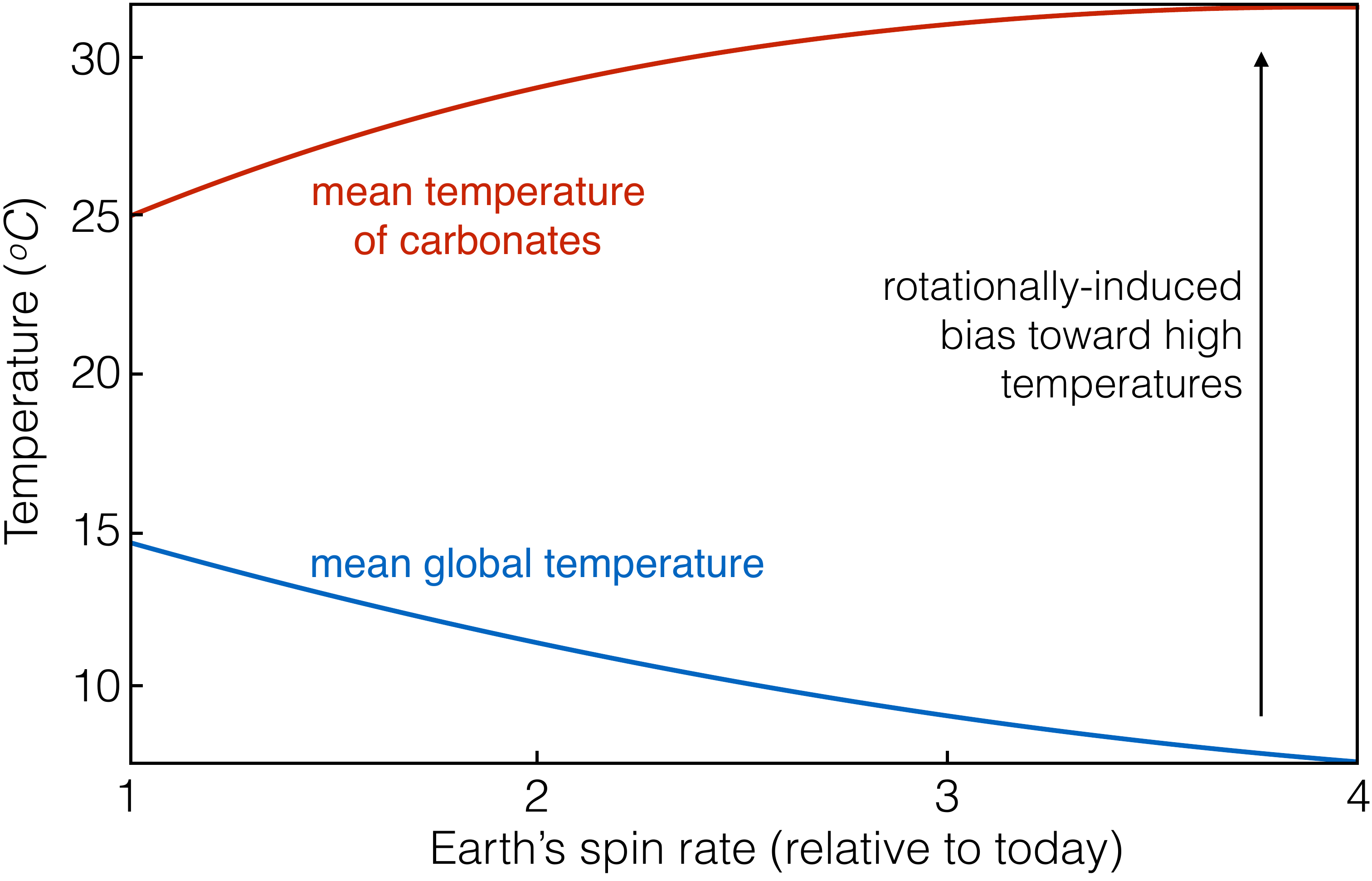}
\caption{The true, global mean temperature $\bar{T}$ (upper, red line) and the temperature that would be inferred from biased carbonate deposition $\bar{T}_{obs}$ (lower, blue line) as a function of rotation rate. We set luminosity equal to Solar for each. Despite the spread of ice sheets leading to a drop in $\bar{T}$, the temperature inferred from the geological record would increase owing to a strong preference toward equatorial deposition.}
\label{Bias}
\end{figure}

Using the definitions above, we computed the dependence of both $\bar{T}$ and $\bar{T}_{obs}$ as a function of $\Omega$, illustrated in Figure~\ref{Bias}. Notably, faster rotation rates led to an enhancement of the mean temperature of precipitated carbonates, despite a reduced global mean temperature. A strong bias remained for all rotation rates, but the magnitude of this bias was amplified as rotation rate increased. Accordingly, care must be taken when inferring a warm Archean climate, simply from an apparently warm signal in sedimentary deposits.

The biases affecting the chert record are less clear, in large part because we know far less about the depositional mechanisms of Precambrian cherts. However, in many cases cherts are likely to have been associated with diagenetic replacements of carbonate rocks \citep{Maliva2005}. The paleoenvironment of many cherts associated with carbonate platforms suggests that they have mechanisms of precipitation wherein evaporative concentration and over saturation may have been important \citep{Fischer2009}. There appear also to have been primary modes of chert deposition as sedimentary particles generated from waters supersaturated with silica \citep{Stefurak2014}. If indeed much of Archean silica deposition was driven by evaporation \citep{Maliva2005,Fischer2009}, the most likely bias in chert formation is at the descending limb of the Hadley cell, which moves equator-ward at faster spin-rates \citep{Kaspi2015}. Given these uncertainties, we did not develop a quantitative model of chert deposition, but to the degree that this is correct, the potential climate bias from observations of chert will plausibly follow a qualitatively similar trend to carbonate deposition.

\subsection{Paleolatitude of shelf environments}

In the previous subsection, we demonstrated that a relatively hotter equator, at higher rotation rates, tends to bias the carbonate record toward warmer environments. However, a key assumption was that all of the Earth's ocean surface was available for carbonate deposition (meaning that in the computation of $\bar{T}_{obs}$, we simply averaged over all latitudes). In reality, geological observations have shown that carbonate deposition during Precambrian time was pronounced in shallow shelf environments \citep{Grotzinger1983,Sumner1996,Ridgwell2003}. A test, in principle, of the suggestion that the Archean record is biased toward the equator, would be to reconstruct the paleolatitudes of sedimentary strata from which the paleoclimate data were generated. 

Limited paleomagnetic evidence from the Barberton Greenstone belt---a critical archive of early Archean paleoclimate data currently---suggests a location close to the paleoequator at $\sim3.5\,$Ga \citep{Kroner1992}. Unfortunately, the paleolatitudes of Archean cratons, and indeed the nature of plate motions at such early times, is debated \citep{Evans2008,Korenaga2013}, reducing confidence in specific reconstructions. 

We took a different approach to examine this using the distribution of paleolatitudes for continental cratons for more recent times, and assume that the same distribution holds for earlier, Archean, times. Our calculation above of $\bar{T}_{obs}$ essentially assumed that over geological time, the distribution of continental cratons is isotropic in latitude. In Figure~\ref{PaleoMag} we constructed a histogram of 45 robust, paleolatitude measurements from \citet{Evans2008} all more ancient than 800\,Ma, and compared it to an isotropic (random) distribution. If no latitude was more likely, the distribution would approximate a cosine---more deposition occurs at lower latitudes in proportion to the greater surface area of the Earth available there. The measured distribution closely approximates that expected from no preferred latitude, validating our computation of $\bar{T}_{obs}$. 

To conclude, the Archean oceans, and global mean temperature more broadly, are likely not as warm as a straightforward read of the archive suggests. The faster-spinning Earth augmented the equatorial temperature relative to the poles, substantially increasing the deposition rate of carbonate at low latitudes relative to the poles. The surviving record, thus, is intrinsically biased toward the warmer parts of the planet and amplified by a climate system that was less efficient at moving heat from the equator. This raises the hypothesis that the poles may have retained cold ice caps while the planet preserved paleoclimate data from the warm equator. 

\begin{figure}
\centering
\includegraphics[trim=0cm 0cm 0cm 0cm, clip=true,width=1\columnwidth]{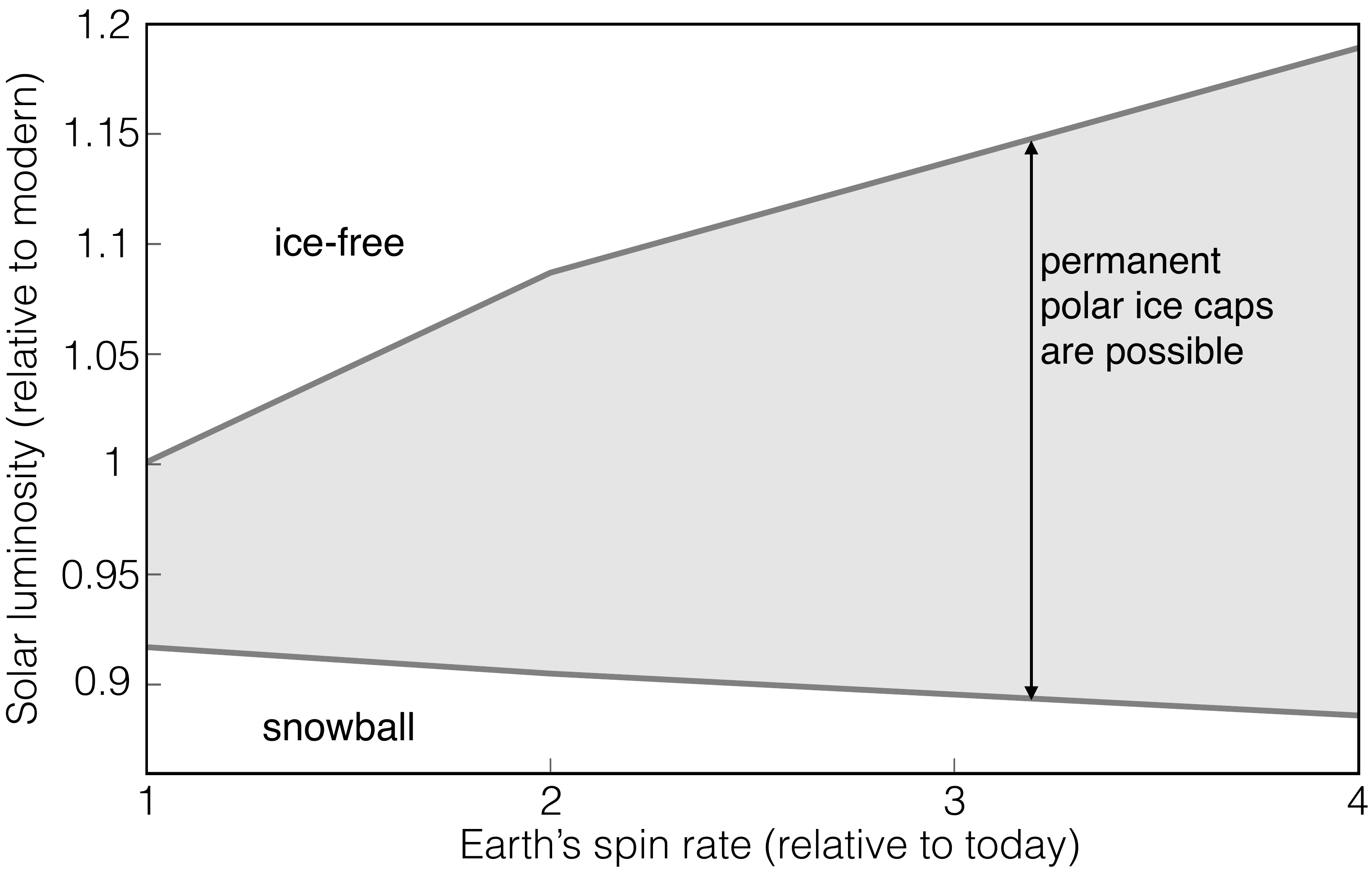}
\caption{The range of Solar luminosities facilitating a partially ice-covered state as a function of Earth's spin rate (see Figure 4). At $\Omega=4\Omega_0$, the increase in Solar luminosity required to deglaciate the planet is many times larger than that required at modern spin rates.}
\label{range}
\end{figure}

\section{Implications for the Faint Young Sun Paradox}

A fundamental aspect of the Archean Earth was that its day was significantly shorter---and it is valuable to frame problems regarding early climate in this context. The enhanced spin rate would have influenced the transport of heat and the distribution of temperatures over the globe. In turn this would have impacted the locus of deposition of chemical sediments. Archean Earth, like the modern, deposited calcium carbonate salts in order to balance weathering fluxes into the oceans. As opposed to the biogenic dominance of such deposition today, the Archean carbonate cycle would largely have been driven by abiotic influences, principally temperature. Combined with the enhanced equatorial temperatures, this leads to a significant, kinetic bias toward deposition of carbonates at low latitudes (Figure~\ref{latitudeOmega}). Given that a substantial fraction of the surviving Archean sedimentary record formed as carbonate, our picture of the Archean world is skewed towards these warmer temperatures.

We also demonstrated that a faster-rotating Earth was generally less sensitive to the ice-albedo feedback. This lower sensitivity makes it substantially more difficult to generate an ice-free planet 3.5 billion years ago. Simply put, the Faint Young Sun paradox is far easier to explain if Earth always maintained ice at the poles than it would be if the planet did not \citep{Wolf2013}. Over the past 500 million years, we have a decent understanding of when the planet was glaciated and when it was not, from a combination of sedimentological and geochemical datasets.

Going further back in time the geological record is sufficiently fragmented that it becomes extremely challenging to discern glacial and ice-free intervals. The gold standard geological observations for early Earth's climate come from glacial deposits (diamictites, ice-rafted debris, striated clasts and pavements)---but these records are inherently sparse. The earliest observations of glacial rocks in the record come from the $\sim3\,$billion-year-old Pongola and Witswatersrand basins of South Africa \citep{Young1998}; yet earlier glacial deposits from $\sim3.5\,$Ga have been described \citep{deWitt2016} though alternative non-glaciogenic interpretations remain possible. Results shown here, however, illustrate just how valuable those sedimentological and geochemical observations are. If further geological investigation finds sufficiently strong evidence of ice-free states of the early Earth, then the faint young Sun paradox becomes exceedingly more difficult to solve. 

In a simple global model, Solar luminosity may be considered as an equivalent to greenhouse gas forcing. Figure~2 in \citet{Ikeda1999} demonstrates a similar result to our Figure~\ref{Hysteresis}, but with $p$CO$_2$ in place of Solar luminosity. A comparison between our result and theirs shows that upper limit of the region facilitating multiple stable states corresponds to $\sim1-2$ orders of magnitude greater $p$CO$_2$ under Archean spin rates than the modern equivalent. In turn, roughly 1 order of magnitude lower $p$CO$_2$ levels are tolerated at faster spins before only a single snowball state is stable.

If an ice-free Earth is required to match the geological record, a resolution to the faint young Sun problem may require revisiting the founding premise behind the whole paradox. I.e., precisely how well do we understand the Solar luminosity over time? If, for example, the Sun's mass had been a few percent higher in the Archean, its higher luminosity would have valuable implications for how we think about the composition of Earth's early atmosphere \citep{Feulner2012}. This idea is promising and remains to be thoroughly tested \citep{Spalding2018}, but is somewhat inconsistent with mass-loss rates observed in Sun-like stars.

The Archean sedimentary record presents us with a valuable, yet limited set of robust windows into its climate, and drivers of habitability. It is critical to thoroughly understand the filters applied at the moment of sedimentation in order to reconstruct the conditions that truly persisted soon after the origin of life. We demonstrated the key role played by rotation rate in sculpting the distribution of temperature, and therefore carbonate deposition, upon the early Earth. These biases yield an apparently warmer Archean than may have existed in reality.

\begin{figure}[h!]
\centering
\includegraphics[trim=0cm 0cm 0cm 0cm, clip=true,width=1\columnwidth]{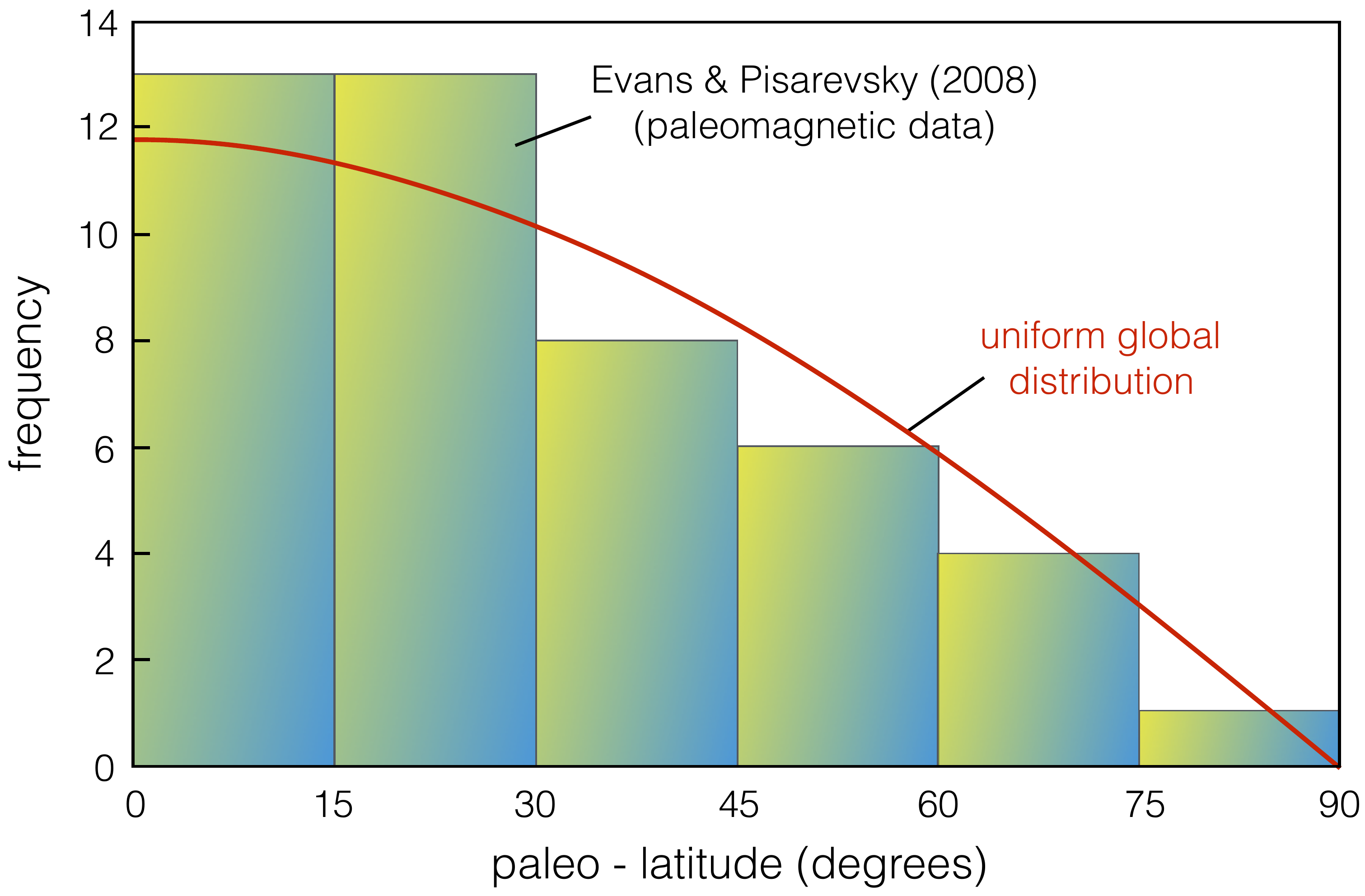}
\caption{The distribution of paleolatitudes older than 800 Ma in a dataset complied by \citet{Evans2008} (bars). If the distribution was uniformly distributed across the globe, the bars and red line should coincide. The data show that low latitude cratons---and their sedimentary basins---are slightly overrepresented.}
\label{PaleoMag}
\end{figure}

\section*{Acknowledgements}
C.S thanks the generous support of the NASA NESSF Graduate Fellowship in Earth and Planetary Science, and of the Heising-Simons Foundation's 51 Pegasi b Postdoctoral Fellowship. W.F. acknowledges support from the Simons Foundation Collaboration on the Origins of Life. Additionally, we thank Konstantin Batygin, Greg Laughlin, Dorian Abbot and Seth Finnegan for illuminating conversations. Finally, we thank several referees for their important contributions to the work. 

%\showacknow{} % Display the acknowledgments section
%\bibliographystyle{apa}
%\bibliography{FYS_References} 

\begin{thebibliography}{}

\bibitem[\protect\astroncite{Abbot et~al.}{2011}]{Abbot2011}
Abbot, D.~S., Voigt, A., and Koll, D. (2011).
\newblock The jormungand global climate state and implications for
  neoproterozoic glaciations.
\newblock {\em Journal of Geophysical Research: Atmospheres}, 116(D18).

\bibitem[\protect\astroncite{Bartlett and Stevenson}{2016}]{Bartlett2016}
Bartlett, B.~C. and Stevenson, D.~J. (2016).
\newblock Analysis of a precambrian resonance-stabilized day length.
\newblock {\em Geophysical Research Letters}, 43(11):5716--5724.

\bibitem[\protect\astroncite{Bills and Ray}{1999}]{Bills1999}
Bills, B.~G. and Ray, R.~D. (1999).
\newblock Lunar orbital evolution: A synthesis of recent results.
\newblock {\em Geophysical Research Letters}, 26(19):3045--3048.

\bibitem[\protect\astroncite{Budyko}{1969}]{Budyko1969}
Budyko, M.~I. (1969).
\newblock The effect of solar radiation variations on the climate of the earth.
\newblock {\em tellus}, 21(5):611--619.

\bibitem[\protect\astroncite{de~Wit and Furnes}{2016}]{deWitt2016}
de~Wit, M.~J. and Furnes, H. (2016).
\newblock 3.5-ga hydrothermal fields and diamictites in the barberton
  greenstone belt paleoarchean crust in cold environments.
\newblock {\em Science advances}, 2(2):e1500368.

\bibitem[\protect\astroncite{Egbert and Ray}{2000}]{Egbert2000}
Egbert, G. and Ray, R. (2000).
\newblock Significant dissipation of tidal energy in the deep ocean inferred
  from satellite altimeter data.
\newblock {\em Nature}, 405(6788):775.

\bibitem[\protect\astroncite{Evans and Pisarevsky}{2008}]{Evans2008}
Evans, D.~A. and Pisarevsky, S.~A. (2008).
\newblock Plate tectonics on early earth? weighing the paleomagnetic evidence.
\newblock {\em When did plate tectonics begin on planet Earth}, 440:249--263.

\bibitem[\protect\astroncite{Feulner}{2012}]{Feulner2012}
Feulner, G. (2012).
\newblock The faint young sun problem.
\newblock {\em Reviews of Geophysics}, 50(2).

\bibitem[\protect\astroncite{Fischer and Knoll}{2009}]{Fischer2009}
Fischer, W.~W. and Knoll, A.~H. (2009).
\newblock An iron shuttle for deepwater silica in late archean and early
  paleoproterozoic iron formation.
\newblock {\em Geological Society of America Bulletin}, 121(1-2):222--235.

\bibitem[\protect\astroncite{Goldblatt et~al.}{2009}]{Goldblatt2009}
Goldblatt, C., Claire, M.~W., Lenton, T.~M., Matthews, A.~J., Watson, A.~J.,
  and Zahnle, K.~J. (2009).
\newblock Nitrogen-enhanced greenhouse warming on early earth.
\newblock {\em Nature Geoscience}, 2(12):891.

\bibitem[\protect\astroncite{Goody}{2007}]{Goody2007}
Goody, R. (2007).
\newblock Maximum entropy production in climate theory.
\newblock {\em Journal of the atmospheric sciences}, 64(7):2735--2739.

\bibitem[\protect\astroncite{Gough}{1981}]{Gough1981}
Gough, D. (1981).
\newblock Solar interior structure and luminosity variations.
\newblock In {\em Physics of Solar Variations}, pages 21--34. Springer.

\bibitem[\protect\astroncite{Grotzinger and Reed}{1983}]{Grotzinger1983}
Grotzinger, J. and Reed, J. (1983).
\newblock Evidence for primary aragonite precipitation, lower proterozoic (1.9
  ga) rocknest dolomite, wopmay orogen, northwest canada.
\newblock {\em Geology}, 11(12):710--713.

\bibitem[\protect\astroncite{Grotzinger and James}{2000}]{Grotzinger2000}
Grotzinger, J.~P. and James, N.~P. (2000).
\newblock Precambrian carbonates: evolution of understanding.

\bibitem[\protect\astroncite{Grotzinger and Kasting}{1993}]{Grotzinger1993}
Grotzinger, J.~P. and Kasting, J.~F. (1993).
\newblock New constraints on precambrian ocean composition.
\newblock {\em The Journal of Geology}, 101(2):235--243.

\bibitem[\protect\astroncite{Higgins et~al.}{2009}]{Higgins2009}
Higgins, J., Fischer, W., and Schrag, D. (2009).
\newblock Oxygenation of the ocean and sediments: consequences for the seafloor
  carbonate factory.
\newblock {\em Earth and Planetary Science Letters}, 284(1-2):25--33.

\bibitem[\protect\astroncite{Hoffman et~al.}{1998}]{Hoffman1998}
Hoffman, P.~F., Kaufman, A.~J., Halverson, G.~P., and Schrag, D.~P. (1998).
\newblock A neoproterozoic snowball earth.
\newblock {\em science}, 281(5381):1342--1346.

\bibitem[\protect\astroncite{Ikeda and Tajika}{1999}]{Ikeda1999}
Ikeda, T. and Tajika, E. (1999).
\newblock A study of the energy balance climate model with co2-dependent
  outgoing radiation: Implication for the glaciation during the cenozoic.
\newblock {\em Geophysical Research Letters}, 26(3):349--352.

\bibitem[\protect\astroncite{Jiang et~al.}{2015}]{Jiang2015}
Jiang, L.-Q., Feely, R.~A., Carter, B.~R., Greeley, D.~J., Gledhill, D.~K., and
  Arzayus, K.~M. (2015).
\newblock Climatological distribution of aragonite saturation state in the
  global oceans.
\newblock {\em Global Biogeochemical Cycles}, 29(10):1656--1673.

\bibitem[\protect\astroncite{Kaspi and Showman}{2015}]{Kaspi2015}
Kaspi, Y. and Showman, A.~P. (2015).
\newblock Atmospheric dynamics of terrestrial exoplanets over a wide range of
  orbital and atmospheric parameters.
\newblock {\em The Astrophysical Journal}, 804(1):60.

\bibitem[\protect\astroncite{Kasting}{2010}]{Kasting2010}
Kasting, J. (2010).
\newblock {\em How to find a habitable planet}.
\newblock Princeton University Press.

\bibitem[\protect\astroncite{Kasting et~al.}{1993}]{Kasting1993}
Kasting, J.~F., Whitmire, D.~P., and Reynolds, R.~T. (1993).
\newblock Habitable zones around main sequence stars.
\newblock {\em Icarus}, 101(1):108--128.

\bibitem[\protect\astroncite{Knauth and Lowe}{2003}]{Knauth2003}
Knauth, L.~P. and Lowe, D.~R. (2003).
\newblock High archean climatic temperature inferred from oxygen isotope
  geochemistry of cherts in the 3.5 ga swaziland supergroup, south africa.
\newblock {\em Geological Society of America Bulletin}, 115(5):566--580.

\bibitem[\protect\astroncite{Korenaga}{2013}]{Korenaga2013}
Korenaga, J. (2013).
\newblock Initiation and evolution of plate tectonics on earth: theories and
  observations.
\newblock {\em Annual Review of Earth and Planetary Sciences}, 41:117--151.

\bibitem[\protect\astroncite{Kr{\"o}ner and Layer}{1992}]{Kroner1992}
Kr{\"o}ner, A. and Layer, P. (1992).
\newblock Crust formation and plate motion in the early archean.
\newblock {\em Science}, 256(5062):1405--1411.

\bibitem[\protect\astroncite{Liu et~al.}{2017}]{Liu2017}
Liu, X., Battisti, D.~S., and Roe, G.~H. (2017).
\newblock The effect of cloud cover on the meridional heat transport: Lessons
  from variable rotation experiments.
\newblock {\em Journal of Climate}, 30(18):7465--7479.

\bibitem[\protect\astroncite{Lorenz et~al.}{2001}]{Lorenz2001}
Lorenz, R.~D., Lunine, J.~I., Withers, P.~G., and McKay, C.~P. (2001).
\newblock Titan, mars and earth: Entropy production by latitudinal heat
  transport.
\newblock {\em Geophysical Research Letters}, 28(3):415--418.

\bibitem[\protect\astroncite{Maliva et~al.}{2005}]{Maliva2005}
Maliva, R.~G., Knoll, A.~H., and Simonson, B.~M. (2005).
\newblock Secular change in the precambrian silica cycle: insights from chert
  petrology.
\newblock {\em Geological Society of America Bulletin}, 117(7-8):835--845.

\bibitem[\protect\astroncite{Meyers and Malinverno}{2018}]{Meyers2018}
Meyers, S.~R. and Malinverno, A. (2018).
\newblock Proterozoic milankovitch cycles and the history of the solar system.
\newblock {\em Proceedings of the National Academy of Sciences},
  115(25):6363--6368.

\bibitem[\protect\astroncite{North}{1975}]{North1975}
North, G.~R. (1975).
\newblock Analytical solution to a simple climate model with diffusive heat
  transport.
\newblock {\em Journal of the Atmospheric Sciences}, 32(7):1301--1307.

\bibitem[\protect\astroncite{Ridgwell et~al.}{2003}]{Ridgwell2003}
Ridgwell, A.~J., Kennedy, M.~J., and Caldeira, K. (2003).
\newblock Carbonate deposition, climate stability, and neoproterozoic ice ages.
\newblock {\em Science}, 302(5646):859--862.

\bibitem[\protect\astroncite{Romanek et~al.}{2011}]{Romanek2011}
Romanek, C.~S., Morse, J.~W., and Grossman, E.~L. (2011).
\newblock Aragonite kinetics in dilute solutions.
\newblock {\em Aquatic geochemistry}, 17(4-5):339.

\bibitem[\protect\astroncite{Sagan and Mullen}{1972}]{Sagan1972}
Sagan, C. and Mullen, G. (1972).
\newblock Earth and mars: evolution of atmospheres and surface temperatures.
\newblock {\em Science}, 177(4043):52--56.

\bibitem[\protect\astroncite{Sellers}{1969}]{Sellers1969}
Sellers, W.~D. (1969).
\newblock A global climatic model based on the energy balance of the
  earth-atmosphere system.
\newblock {\em Journal of Applied Meteorology}, 8(3):392--400.

\bibitem[\protect\astroncite{Som et~al.}{2016}]{Som2016}
Som, S.~M., Buick, R., Hagadorn, J.~W., Blake, T.~S., Perreault, J.~M.,
  Harnmeijer, J.~P., and Catling, D.~C. (2016).
\newblock Earth's air pressure 2.7 billion years ago constrained to less than
  half of modern levels.
\newblock {\em Nature Geoscience}, 9(6):448.

\bibitem[\protect\astroncite{Spalding et~al.}{2018}]{Spalding2018}
Spalding, C., Fischer, W.~W., and Laughlin, G. (2018).
\newblock An orbital window into the ancient sun's mass.
\newblock {\em The Astrophysical Journal Letters}, 869(1):L19.

\bibitem[\protect\astroncite{Stefurak et~al.}{2014}]{Stefurak2014}
Stefurak, E.~J., Lowe, D.~R., Zentner, D., and Fischer, W.~W. (2014).
\newblock Primary silica granules-a new mode of paleoarchean sedimentation.
\newblock {\em Geology}, 42(4):283--286.

\bibitem[\protect\astroncite{Sumner and Grotzinger}{1996}]{Sumner1996}
Sumner, D.~Y. and Grotzinger, J.~P. (1996).
\newblock Were kinetics of archean calcium carbonate precipitation related to
  oxygen concentration?
\newblock {\em Geology}, 24(2):119--122.

\bibitem[\protect\astroncite{Touma and Wisdom}{1994}]{Touma1994}
Touma, J. and Wisdom, J. (1994).
\newblock Evolution of the earth-moon system.
\newblock {\em The Astronomical Journal}, 108:1943--1961.

\bibitem[\protect\astroncite{Vallis and Farneti}{2009}]{Vallis2009}
Vallis, G.~K. and Farneti, R. (2009).
\newblock Meridional energy transport in the coupled atmosphere--ocean system:
  Scaling and numerical experiments.
\newblock {\em Quarterly Journal of the Royal Meteorological Society},
  135(644):1643--1660.

\bibitem[\protect\astroncite{Walker et~al.}{1981}]{Walker1981}
Walker, J.~C., Hays, P., and Kasting, J.~F. (1981).
\newblock A negative feedback mechanism for the long-term stabilization of
  earth's surface temperature.
\newblock {\em Journal of Geophysical Research: Oceans}, 86(C10):9776--9782.

\bibitem[\protect\astroncite{Webb}{1982}]{Webb1982}
Webb, D. (1982).
\newblock Tides and the evolution of the earth-moon system.
\newblock {\em Geophysical Journal of the Royal Astronomical Society},
  70(1):261--271.

\bibitem[\protect\astroncite{Williams and Kasting}{1997}]{Williams1997}
Williams, D.~M. and Kasting, J.~F. (1997).
\newblock Habitable planets with high obliquities.
\newblock {\em Icarus}, 129(1):254--267.

\bibitem[\protect\astroncite{Williams}{2000}]{Williams2000}
Williams, G.~E. (2000).
\newblock Geological constraints on the precambrian history of earth's rotation
  and the moon's orbit.
\newblock {\em Reviews of Geophysics}, 38(1):37--59.

\bibitem[\protect\astroncite{Wolf and Toon}{2013}]{Wolf2013}
Wolf, E. and Toon, O. (2013).
\newblock Hospitable archean climates simulated by a general circulation model.
\newblock {\em Astrobiology}, 13(7):656--673.

\bibitem[\protect\astroncite{Wordsworth and
  Pierrehumbert}{2013}]{Wordsworth2013}
Wordsworth, R. and Pierrehumbert, R. (2013).
\newblock Hydrogen-nitrogen greenhouse warming in earth's early atmosphere.
\newblock {\em science}, 339(6115):64--67.

\bibitem[\protect\astroncite{Yang et~al.}{2012}]{Yang2012}
Yang, J., Peltier, W., and Hu, Y. (2012).
\newblock The initiation of modern soft and hard snowball earth climates in
  ccsm4.
\newblock {\em Climate of the Past}, 8(3):907--918.

\bibitem[\protect\astroncite{Young et~al.}{1998}]{Young1998}
Young, G.~M., Brunn, V.~V., Gold, D.~J., and Minter, W. (1998).
\newblock Earth's oldest reported glaciation: physical and chemical evidence
  from the archean mozaan group (~ 2.9 ga) of south africa.
\newblock {\em The Journal of geology}, 106(5):523--538.

\bibitem[\protect\astroncite{Zahnle and Walker}{1987}]{Zahnle1987}
Zahnle, K. and Walker, J.~C. (1987).
\newblock A constant daylength during the precambrian era?
\newblock {\em Precambrian Research}, 37(2):95--105.

\bibitem[\protect\astroncite{Zeebe and Wolf-Gladrow}{2001}]{Zeebe2001}
Zeebe, R.~E. and Wolf-Gladrow, D.~A. (2001).
\newblock {\em CO2 in seawater: equilibrium, kinetics, isotopes}.
\newblock Number~65. Gulf Professional Publishing.

\end{thebibliography}

\appendix
\section{Solution to the energy balance model}

In this appendix, we provided the derivation of the solution to the Budyko-Sellers energy-balance climate model. The equation describing the steady-state meridional temperature distribution $T$, written in terms of a new variable $x\equiv \sin(\lambda)$, takes the form \citep{North1975}

\begin{linenomath*}
\begin{align}\label{Legendre}
\underbrace{-D \frac{d}{dx}(1-x^2)\frac{dT}{dx}}_{\textrm{diffusive transport}}=\underbrace{Q a(x,x_s)S(x)}_{\textrm{Solar insolation}}- \underbrace{I(T,p\textrm{CO}_2)}_{\textrm{re-emitted heat}},
\end{align}
\end{linenomath*}

where the coefficient 

\begin{linenomath*}
\begin{align}
D\equiv \frac{c_p \Phi}{g R^2}K_{\textrm{eff}},
\end{align}
\end{linenomath*}

and $Q$ is the Solar constant divided by 4. In the above, $c_p=10^3$J\,kg$^{-1}$K$^{-1}$ is the specific heat of air, $\Phi=1$\,bar is the surface air pressure, $g=10$\,m\,s$^{-2}$ is the acceleration due to gravity and $R=6400$\,km is the Earth's radius. The Solar insolation function $S(x)$ is defined as 
\begin{linenomath*}
\begin{align}
S(x)&=1+S_2 P_2(x)\nonumber\\
&=1+S_2 \frac{1}{2}(3x^2-1)
\end{align}
\end{linenomath*}

where $P_2(x)$ is a degree 2 Legendre Polynomial. The function $a(x,x_s)$ yields the fraction of incoming Solar radiation that is absorbed (one minus the albedo) as a function of latitude. The ice-albedo feedback is built into $a(x,x_s)$; defined as
\begin{align}
a(x,x_s)
\begin{cases}
    a_0,& \text{if } x\leq x_s\\
    a_1,              & \text{otherwise}
\end{cases}
\end{align}
where $a_0>a_1$, as ice is more reflective than ocean. 

The outgoing radiation is prescribed through the function 
\begin{linenomath*}
\begin{align}
I(T)=A+B T(x)
\end{align}
\end{linenomath*}
where $A$ and $B$ are constants to be determined by tuning to the modern climate. 
Such linearity allows us to replace $T(x)$ with $I(T)/B$ as the dependent variable \citep{North1975}. Carrying out the substitution and defining 
\begin{linenomath*}
\begin{align}
D'\equiv \frac{D}{B}
\end{align}
\end{linenomath*}

we may formulate the problem into an ordinary differential equation satisfying
\begin{linenomath*}
\begin{align}
\frac{d}{dx}(1-x^2)\frac{d I(x)}{dx}-\frac{1}{D'}I(x)=-\frac{Q}{D'}S(x)a(x,x_s).
\end{align}
\end{linenomath*}

In order to solve the differential equation above, we separated the solution into two regimes; one at low latitudes where the albedo is smaller and a second at high latitudes where the albedo is that of ice. For this part of the calculation, we ignored the effect of spin rate upon albedo, and included only the effect of rotation rate upon diffusivity.  

In each region ($i=0$ for low latitudes and $i=1$ for high latitudes), the general solution may be written \citep{North1975}
\begin{linenomath*}
\begin{align}
I_i&(x)=\mathcal{A}_i P_l (x)+\mathcal{B}_i Q_l (x)\nonumber\\
&+Q a_i \bigg[1+\frac{S_2 P_2(x)}{6 D' +1}\bigg],\,\,\,\,i=0,\,1
\end{align}
\end{linenomath*}
where $\mathcal{A}_i$ and $\mathcal{B}_i$ are constants to be determined by satisfying boundary conditions, and the parameter $l$ is defined as
\begin{linenomath*}
\begin{align}
D'=-\frac{1}{l(1+l)}.
\end{align}
\end{linenomath*}
The boundary conditions require zero heat flux at the equator and pole. They also require continuity of gradient at $x=x_s$ and at this point $I_0=I_1=I_s$, which was computed using a predefined critical temperature for permanent ice caps ($T=T_s\equiv-10^\circ$C).

\end{document}